\documentclass[]{aa}
\usepackage{txfonts}
\usepackage{graphicx}
\usepackage{natbib}

\newcommand{\kms}{$\rm km\,s^{-1}$}
\newcommand{\am}{$^{\prime}$}

\newcommand{\hi}{\ion{H}{I}}
\newcommand{\hii}{\ion{H}{II}}

\newcommand{\gs}{GS\,100--02--41}

\begin{document}
\title{GS100-02-41: a new large \hi\, shell in the outer part of the Galaxy.}
\author{L. A. Suad\inst{1}
\and S. Cichowolski\inst{2} \and   E. M. Arnal \inst{1,3}
\and J. C. Testori\inst{1} }

\institute{Instituto Argentino de Radioastronom\'{\i}a (IAR)(CCT - La PLata), CC 5,
  1894, Villa Elisa, Argentina. \and 
   Instituto de Astronom\'{\i}a y F\'{\i}sica del
  Espacio (IAFE), CC 67, Suc. 28, 1428 Buenos Aires, Argentina.
\and Facultad de Ciencias
  Astron\'omicas y Geof\'{\i}sicas, Universidad Nacional de La Plata, Argentina.}

\date{received  /accepted  }

\abstract 
{Massive stars have a profound effect on the surrounding interstellar medium. They ionize and heat the neutral gas, and due to 
their strong 
winds, they swept  the gas up forming large \hi\ shells. In this way, they generate a dense shell where the  physical 
conditions for the 
formation of new stars are given.}
{The aim of this study is to analyze the origin and evolution of the large \hi\ shell \gs\, and its role in triggering star forming processes.}{To characterize the shell and its environs, 
we carry out a multi-wavelength study. We analyze 
 the \hi\ 21 cm line, 
 the radio continuum, and infrared emission distributions.  }{The analysis of the \hi\ data shows an expanding  shell 
structure  centred at ($l, b$) = (100\fdg6, --2\fdg04) in the 
velocity range from  --29 to --51.7  \kms. 
Taking into account non circular motions, we infer for \gs\, a  kinematical distance of 2.8 $\pm$ 0.6 kpc. 
Several massive stars belonging to Cep\,OB1 
are located in projection within the  large \hi\, shell boundaries. The analysis of the radio continuum and infrared data reveal that there is no continuum 
counterpart of the \hi\ shell. On the other hand, three slightly extended radio continuum sources are observed in projection onto the dense \hi\ 
shell. From their 
flux density determinations we infer that they are thermal in nature. An analysis of the \hi\ emission distribution in the environs 
of  these sources shows, for each of them,  a region of low emissivity  having a good morphological correlation with the
 ionized gas in a velocity range similar to the one where \gs\, is detected.}
{Based on an  energetic analysis, we conclude that the origin of \gs\, could have been mainly due to the action of the Cep\,OB1 massive stars 
located inside the \hi\ shell. The obtained age difference between  the \hi\ shell and the \hii\ regions, together with their relative 
location, led us to conclude that the ionizing stars could have been created as a consequence of the shell evolution. }


\maketitle

\titlerunning{}
\authorrunning{L. A. Suad et al.}

\section{Introduction}

The presence in the interstellar medium (ISM) of the Milky Way, when viewed in the $\lambda\sim$ 21 cm line emission of the 
neutral hydrogen (\hi), of giant
structures having linear dimensions of a few hundred parsecs in diameter is a widely
known and well observed phenomena first noticed by
\citet{hei79}. These structures are usually detected as huge shells or arc-like
features of enhanced  \hi\, emission
surrounding regions of low \hi\,  emissivity, receiving the
generic name of \hi\, supershells. These features may even be the dominant structure in the interstellar medium, taking up a large fraction of the volume of the galactic disk. The \hi\, structures could be also observed at infrared wavelengths.
Based on the 60 and 100 $\mu$m IRAS databases, \citet{kon07} have performed an all-sky survey of loop- and arc-like structures.

Similar structures have also been
observed in nearby spiral galaxies  \citep{sta07,cha11}. 
In the Milky Way, these structures were initially catalogued
by  \citet{hei79,hei84}. Though a large number of \hi\, features
likely to be classified as either large \hi\, shells or \hi\,
supershells have been catalogued in the outer (90$^\circ \leq l
 \leq$ 270$^\circ$) part of the Galaxy \citep{ehl05}, only a small number of them have been studied in some detail.
The later \citep{jun96, sti01, uya02, mcc02, 
caz03, arn07, cic11} have galactocentric
distances ranging from 9.7 to 16.6 kpc, diameters from 120 to $\sim$ 840 pc,
expansion velocities between $\sim$10 and $\sim$ 20 \kms, and kinetic energies
from $\sim$ 1 $\times$ 10$^{50}$ up to $\sim$ 6 $\times$ 10$^{51}$ erg.
Among them only two (GS\,305+01-24 and the feature studied by \citet{caz03})  have an OB-association as their
 likely powering source, and other three (GSH\,91.5+2--114, GS\,234--02 and GS\,263--02+45) show evidence of having induced the formation of new generation of stars.

The general consensus is that those structures whose kinetic energy is
of the order of, or less than, a few times 10$^{51}$ erg, very likely may 
have been created by the joint action of stellar winds and supernova
 explosions. A large
number of examples are reported in the literature \citep[e.g.][]{uya02,arn07,cic11}. On the other hand, for expanding \hi\, structures having kinetic
energies in excess $\sim 10^{52}$ erg, termed {\it supershells}, 
the above mechanism may not be adequate to create them because one would
need a stellar grouping (either an open cluster or an OB-association) with
many more stars than the average found in the Milky Way. In these cases
alternative mechanisms like the infalling of high velocity clouds \citep{ten81}
or gamma-ray bursts \citep{per00} may be at work.

Along its expansion, these
structures (either a shell or a supershell) may became gravitationally
unstable, forming clouds that later on may lead to the formation of stars along
the periphery of these \hi\, structures, or else the expanding structure
may hit and compress  pre-existing {\rm ISM} molecular clouds from one side.
 During this process a high density perturbation may move into the molecular
 clouds, which may eventually collapse into denser cores in which star
formation may occur. A thorough review of observations and theory related to
 trigger star formation is given by \citet{elm98}.

A new large scale study aim at detecting  in the outer part of the galaxy structures likely to be either large \hi\, shells or supershells is being carried out by one of the authors (L. A. Suad) as
part of her PhD Thesis.  

In this paper we analyze a new large \hi\, shell  observed at (\textit{l, b}) $\sim$ ($100^\circ$, $-2^\circ$),
with the purpose of elucidating both its origin and  its interaction with the surrounding ISM. We also
look for signs of recent star formation activity likely to be related to this shell.

\section{Observations}

Low resolution \hi\, data were retrieved from the
Leiden-Argentine-Bonn (LAB) survey \citep{kal05}. This database is well
suited for a study of large scale structures due to its angular
resolution. The entire database has
been corrected for stray radiation.  High angular resolution \hi\, data, covering most of the structure under
study, were obtained from the Canadian Galactic Plane Survey 
\citep[CGPS,][]{tay03}.  Besides the high resolution \hi\, data, the CGPS also
provides high resolution continuum data at 408 and 1420 MHz
\citep{lan00}.  Continuum data at 2695 MHz
\citep{rei84, rei90, fur90} were also used in this study.

Infrared data from the Midcourse Space Experiment (MSX) \citep{pri01}
 were obtained from the Infrared Science
Archive \footnote{The NASA/IPAC Infrared Science Archive is operated
by the Jet Propulsion Laboratory, California Institute of
Technology, under contract with the Nasa Aeronautics and Space
Administration (http://irsa.ipac.caltech.edu)}.  Infrared images
from the Improved Reprocessing of the IRAS Survey (IRIS) \citep{miv05} were
retrieved from the SkyView
homepage \footnote{http://skyview.gsfc.nasa.gov/}.

Carbon monoxide data ($^{12}$CO (1$\rightarrow$0)) for a small area centered at ($l, b$) 
$\sim$ ($100\fdg7, -0\fdg5$) were kindly made available  to us by Dr. Christ Brunt, while for $l$ $\geq 102\fdg08$  $^{12}$CO 
data were retrieved from the CGPS database. These data were obtained using the 14-m dish of the  Five College Radio Astronomy 
Observatory (FCRAO).

Table \ref{tabla-observations} summarizes the most relevant observational parameters.

The use of this multi-wavelength approach allows us to probe the
different components (neutral and
ionized gas and  dust) of the ISM.

\begin{table}
\caption{Observational parameters.} 
\label{tabla-observations}

\centering
    \begin{tabular}{l c}
 \hline\hline
\textbf{LAB \hi\, data} &   \\
Angular resolution & 34\arcmin \\
Velocity resolution & 1.3 \kms \\
Velocity coverage & --450 to 400 \kms \\
\textbf{CGPS \hi\, data} &   \\
Angular resolution & $1\farcm2 \times 1\arcmin $ \\
Velocity resolution & 1.3 \kms \\
Velocity coverage & --164 to 58 \kms\\
\textbf{Radio continuum} & \\
Angular resolution \textbf{(408 MHz)}& $3\farcm5 \times 4\farcm6$ \\
Angular resolution \textbf{(1420 MHz)}& $1\farcm2  \times 1\arcmin$ \\
Angular resolution \textbf{(2695 MHz)}& 4\farcm3 \\
\textbf{CO data} & \\
Angular resolution & $\sim$ 1\arcmin\\
Velocity resolution & 0.824 \kms \\
\textbf{Infrared data} & \\
Angular resolution \textbf{(MSX)} & 18\farcs4\\
Angular resolution \textbf{(IRIS)}& 3\farcm8 -- 4\farcm3\\
Angular resolution \textbf{(HIRES)}& 0\farcm5 -- 2\arcmin\\
\hline
\end{tabular}
\end{table}

\section{\gs.}

\subsection{Neutral hydrogen data.}\label{hi}

We consider that a given  \hi\, structure may be classified as a shell if the following criteria are fullfilled:

\begin{enumerate}

\item It must have a well defined lower brightness temperature
  surrounded (partially or completely) by regions of higher
 temperature. 

\item  The \hi\, minimum must be observable in at least  5
 consecutive velocity channels.

\item It must have a minimum angular size of $2^\circ$. This condition
  is set bearing in mind the angular resolution of the LAB survey, to assure the structure under study is, angularly speaking, 
fully resolved.

\item  At the kinematic distance of the structure, its linear size must exceed 200 pc.

 \end{enumerate}

\begin{figure*}
\resizebox{\hsize}{!}{\includegraphics[width=17cm]{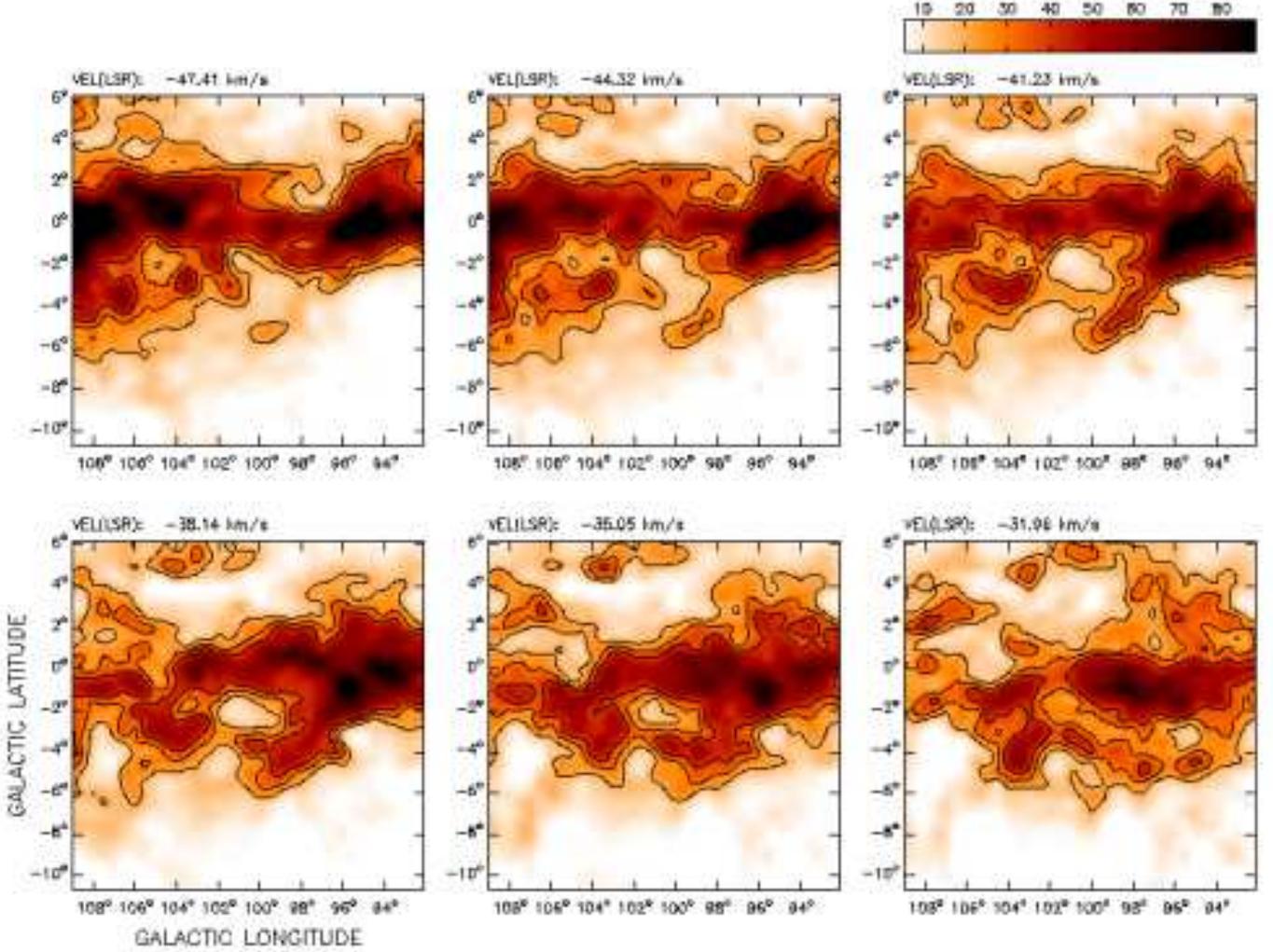}}
   \caption{ LAB \hi\, emission distribution averaged in the velocity range
     from $-47.41$ to $-31.96$ \kms. Each image is an average of three
     velocity channel maps. Contour levels are at 20, 30, and 40 K.
     The LSR central velocity of each image is indicated in its top
     left corner. The color bar-code shows the temperature scale in kelvins for all the images.}
\label{9mosaicos}
\end{figure*}

Relying on the criteria described above, we are constructing  a large \hi\, shells and supershells catalogue in the outer part of the Galaxy
 (Suad et al. in preparation). As part of this catalogue, we have found an \hi\, structure located at 
($l, b$, $V$) = ($100\fdg6$, $-2\fdg3$, --41 \kms). Following the standard nomenclature, this shell is labelled as \gs\,.

To illustrate the main observational findings, in Fig. \ref{9mosaicos} a mosaic of six \hi\, images  covering the velocity 
range form $-47.41$ to $-31.96$  \kms\,  is shown (all velocities in this paper are referred to the LSR). Each image is an 
average of three individual line channel maps and covers a velocity interval of about 3.09 \kms.

The upper left panel shows the \hi\, brightness temperature distribution at $-47.41$ \kms. Besides the \hi\, emission arising 
from the galactic plane, quite a few regions  of low emissivity are  clearly identified. 
These regions appear  as a local minimum surrounded along most of its perimeter by regions of higher brightness temperature, as for example 
 the structure located at  ($l, b$) = (101\fdg0, --2\fdg0). 
As  we move towards  more positive velocities, this structure becomes a well defined \hi\, minimum 
within the overall galactic \hi\, emission. This minimum achieves its maximum angular extent at $-41.2$ \kms\, 
and is barely  visible at $-32$ \kms\, (lower right panel).

\begin{figure}
\resizebox{\hsize}{!}{\includegraphics[width=8cm]{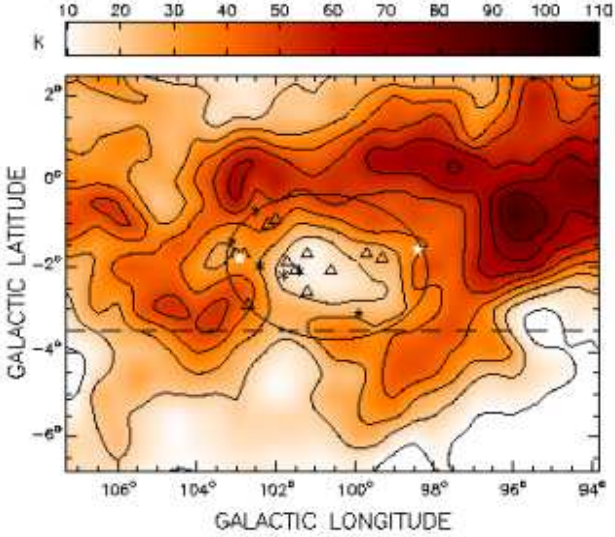}}
   \caption{ LAB \hi\, emission distribution averaged  in the velocity range from --42.26 to --34.02 \kms. 
 Contour levels are from 10 to 90 in steps of 10 K. The area  
below the dashed line corresponds to the region where CGPS data are not available. The  symbols indicate the location of the 
star members of Cep\,OB1 lying inside the ellipse (see Section \ref{originob}). White star symbols correspond to O-type main sequence stars. 
Asterisks and triangles indicate evolved stars that were O-type and B-type stars during the main sequence phase, respectively.}
\label{100-2-37ProV}
\end{figure}

An average of the  \hi\ emission distribution in the velocity range from $-42.26$ to $-34.02$ \kms\, 
is shown in  Fig. \ref{100-2-37ProV}, where a huge \hi\, structure having dimensions of $\sim\, 6^\circ \times 3^\circ$ 
 ($\Delta l$ $\times$ $\Delta b$) is easily recognizable. 

To estimate the main parameters of this large \hi\, shell,
we have characterized the ellipse that best fit it using a least square method. The points used to fit the ellipse correspond to the local maxima around the cavity. To select these points the CGPS \hi\, data were used.
The parameters derived from this fit are: the symmetry centre of
the ellipsoidal \hi\, distribution (\emph{l$_0$, b$_0$}), the length of both the semi-major
(\emph{a}) and semi-minor (\emph{b}) axes of the ellipse, and the
inclination angle ($\theta$) between the major axis and the galactic
longitude axis. This angle is positive towards the north galactic
pole. They are given in Table \ref{tabla1}.

\begin{figure}
\centering
\includegraphics[width=7cm]{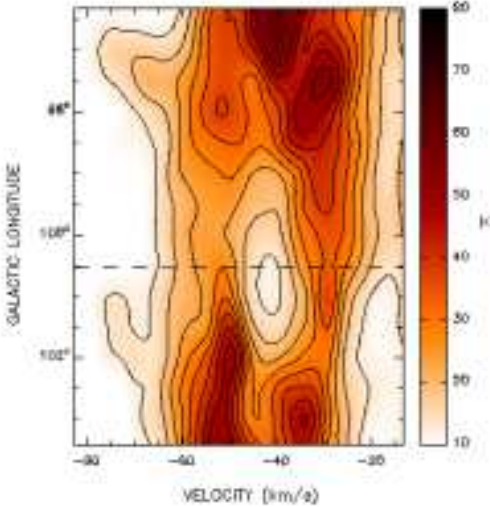}
\caption{ \textbf{LAB}  \hi\, emission
     distribution averaged in the latitude  range from --3\fdg0 to
     --1\fdg3. 
     Contour levels are from 13 to 53 K in steps of 5 K. The dashed line indicates the location where the cross-cut shown 
in Fig. \ref{corte-l=100} was performed.}
\label{v-b}
\end{figure}

\begin{figure}
\centering
\includegraphics[width=7cm]{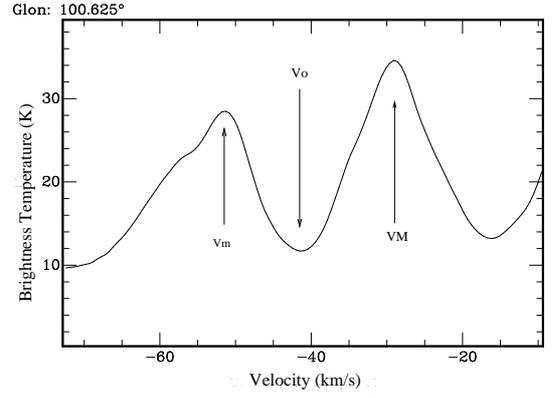}
\caption{\hi\, cross-cut at l=100$\fdg$6 obtained from Fig. \ref{v-b} . The approaching ($V_\mathrm{m}$), 
 systemic ($V_0$) and receding ($V_\mathrm{M}$) velocities are indicated.}
\label{corte-l=100}
\end{figure}

Under the assumption of a symmetric expansion, an ellipsoidal \hi\,
feature having a central velocity $V_0$ and an expansion velocity
$V_{e}$ should depict, in a position-position diagram, an ellipse-like
pattern when observed at different radial velocities. At $V_0$ the
ellipse of \hi\, emission attains its maximum dimensions, while at
extreme velocities (either approaching ($V_\mathrm{m}\,= \, V_0 \, -
\, V_\mathrm{e}$) or receding ($V_\mathrm{M}\,= \, V_0 \, + \,
V_\mathrm{e}$)) the hydrogen emission should look like an "ovoidal"
patch of emission. At intermediate velocities the dimension of the
\hi\, ellipse shrinks as $V_\mathrm{M}\,$ (or $V_\mathrm{m}\,$) is
approached. The expansion velocity is estimated as half of the total
velocity range ($V_{e} = 0.5\, (|V_\mathrm{M}\, - \, V_\mathrm{m}|)$)
covered by the \hi\, emission related to the feature.  This method
always provides a lower limit to $V_{e}$, because \hi\, emission
arising from those regions having radial velocities close to either
the maximum approaching ($V_\mathrm{m}$) or receding ($V_\mathrm{M}$)
cap are usually difficult to disentangle from the overall galactic
\hi\, emission.  Another way to determine $ V_\mathrm{e}$ without the
above drawback is to use the velocity-position diagrams. In
Fig. \ref{v-b} a radial velocity versus galactic longitude diagram is
shown.  This image shows a different view of \gs\,.  The inner region
of the elliptical \hi\, feature shown in Fig. \ref{100-2-37ProV},
corresponds to the low emissivity region seen at ($V$, $l$) = ($-41$
\kms, $100\fdg6$). The strong peaks of \hi\, emission seen at $\sim
-29$ \kms\, and $\sim -51$ \kms\, arise from the \hi\, associated with
the expanding walls of neutral gas related to \gs.  Figure \ref{corte-l=100}
shows a cross-cut of Fig. \ref{v-b} along $l=100\fdg6$. By making a Gaussian fit to the \hi\, emission peaks,
$V_\mathrm{m}$ and $V_\mathrm{M}$ can be derived, along with the value
of $V_0$, that corresponds to the velocity between $V_\mathrm{m}$
and $V_\mathrm{M}$ where the minimum value of $T_b$ is observed.  We
obtain $V_0$ = $-41 \pm 2$ \kms and $V_\mathrm{e}$ = $11 \pm 2$ \kms.

It is well known that  non-circular motions on a large scale are present in the Perseus spiral arm \citep{bra93}, making the galactic rotation 
model  not suitable for this region of the Galaxy.
 Based on the observed velocity field derived by \citet{bra93}, the systemic velocity of \gs\, can be translated to a kinematic distance 
of 2.8 $\pm$ 0.6 kpc. We shall adopt this distance for the \hi\, shell.

Under the assumption that the \hi\, emission is optically thin and following the procedure described by \citet{pin98}, the total neutral hydrogen mass of a structure located at a 
distance \textit{d} (kpc) that subtends a solid angle $\Omega$  (square arc-min) is given by

$$ M_\mathrm{HI}(M_\odot)=1.3\times10^{-3} \, \emph{d}^2 \, \Delta V \,
\Delta\emph{T}_\mathrm{B} \, \Omega $$

\noindent where $\Delta$V is the velocity interval over which the structure is
detected, expressed in \kms\, and $\Delta T_\mathrm{B}$ (K) is the mean brightness temperature defined as 
$\Delta T_\mathrm{B}=\mid
\emph{T}_\mathrm{sh}-\emph{T}_\mathrm{bg} \mid$, where
\emph{T}$_\mathrm{sh}$  refers to the mean brightness temperature of
the \hi\, shell, and \emph{T}$_\mathrm{bg} $ corresponds to the temperature of
the contour level defining the outer border of the \hi\, shell.  
The latter represents the temperature of the  surrounding galactic \hi\, emission gas. 
For this structure we estimated \emph{T}$_\mathrm{sh} =  33.5 \pm 1.6$ K, \emph{T}$_\mathrm{bg} = 21.5 \pm 1.5$ K, $\Delta$V $= 22 \pm 2$ \kms and $\Omega = 4.2 \times 10^4$ arcmin$^2$.
Adopting solar abundances, the total gaseous mass of \gs\, is
$ \emph{M}_\mathrm{t}(M_\odot)=1.34 \, M_\mathrm{HI} $ (see Table \ref{tabla1}).

Assuming that the mass is uniformly distributed within the structure's volume, we obtain  
the gas number density ($n_{sh}$) of the swept up gas. 
The ambient density ($n_0$) of the medium into which the large \hi\, shell
is evolving is derived by uniformly distributing the shell mass
($M_\mathrm{t}$) over the volume swept up by the structure. Assuming this mass
to be distributed into an ovoidal volume whose semi-major axis is
\emph{a} and the other two dimensions are set equal to \emph{b}, then

$$ n_0= 10 \frac {M_\mathrm{t}} {a \, b^2}\,\,  (\mathrm{cm^{-3}}) $$

\noindent where $M_\mathrm{t}$ is given in M$_\odot$, and \emph{a} and \emph{b} in parsecs.

A rough estimate of the age of the shell can be obtained using a simple model to describe the expansion 
of a shell created by a continuous injection of mechanical energy  or by a supernova explosion. 
In this way, the dynamical age of \gs\, can be estimated as $t_{\rm dyn} = \alpha$\,R / $V_e$  \citep{wea77}, where R is the radius of the shell 
($R = \sqrt{ a\, b}$) ,
and $\alpha = 0.25$  for a radiative supernova remnant (SNR) or  $\alpha = 0.6$ for a stellar wind shell.

Another important parameter that characterizes the shells is the kinetic energy, which is given 
by $\emph{E}_\mathrm{k}=0.5 \,  \emph{M}_\mathrm{t} \, V_{e}^2 $. All the relevant parameters of \gs\, 
are given in Table \ref{tabla1}.

\begin {table}
\caption{\gs\, main parameters.} 
\label{tabla1}
\centering
    \begin{tabular}{l c }
 \hline\hline
  Parameter &  Value\\
\hline
Distance (kpc) &  2.8 $\pm$ 0.6 \\
($l_0$, $b_0$)  &  (100\fdg6,  --2\fdg04)\\
a  &2\fdg56\\
b  &1\fdg69\\
$\theta$   & --5\fdg7\\
a (pc)& 125$\pm$ 25\\
b (pc)& 83 $\pm$ 17\\
$V_0$ (\kms) &   --41 $\pm$ 2\\
$V_{e}$ (\kms) &  11 $\pm$ 2\\
M$_\mathrm{t}$ (M$_\odot$) & (1.5$\pm$ 0.7) $\times 10^5$  \\
$n_{sh}$  (cm$^{-3}$) &   2.5 $\pm$ 0.4\\
$n_0$ (cm$^{-3}$) &  1.7 $\pm$ 0.4\\
$\emph{E}_\mathrm{k}$ (erg) &  $ (1.8 \pm 0.8)\times 10^{50} $\\
t$_\mathrm{dyn}$ ($\alpha=0.25$) (Myr) & 2.3 $\pm$ 1.1\\
t$_\mathrm{dyn}$ ($\alpha=0.6$) (Myr) &     5.5 $\pm$ 2.7\\

\hline

 \end{tabular}
\end{table}

\begin{figure}
\resizebox{\hsize}{!}{\includegraphics[width=8cm]{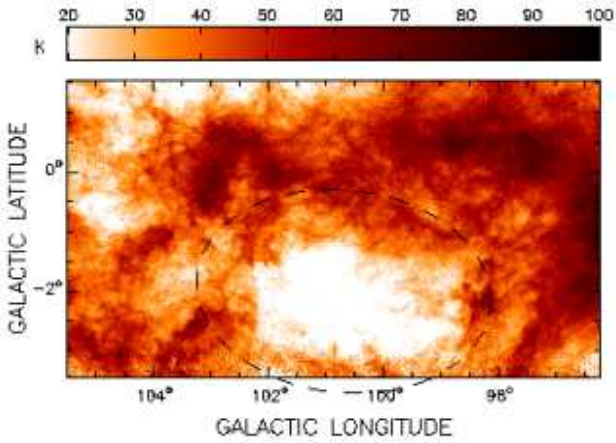}}
   \caption{CGPS averaged \hi\, emission distribution in the velocity range from --42.7 to --34.4 \kms. The dashed ellipse corresponds to the ellipse shown in Fig. \ref{100-2-37ProV}.}
\label{100-2-37ProV-CGPS}
\end{figure}

 Even though the CGPS data do not cover the whole area of \gs, for completeness in Fig. \ref{100-2-37ProV-CGPS} the CGPS data averaged in the same velocity range than the LAB data are shown. A comparison with Fig. \ref{100-2-37ProV} clearly shows that the large-scale structures are observed in both data sets and, as expected, a lot of small scale structures are observed in the high resolution image.
The dashed ellipse shown in Fig. \ref{100-2-37ProV-CGPS} is the one obtained form the fitting of the low resolution data  (see Table \ref{tabla1}).

\subsection{Radio continuum and infrared data.}\label{cont-ir}

\begin{figure}
\includegraphics[width=8cm]{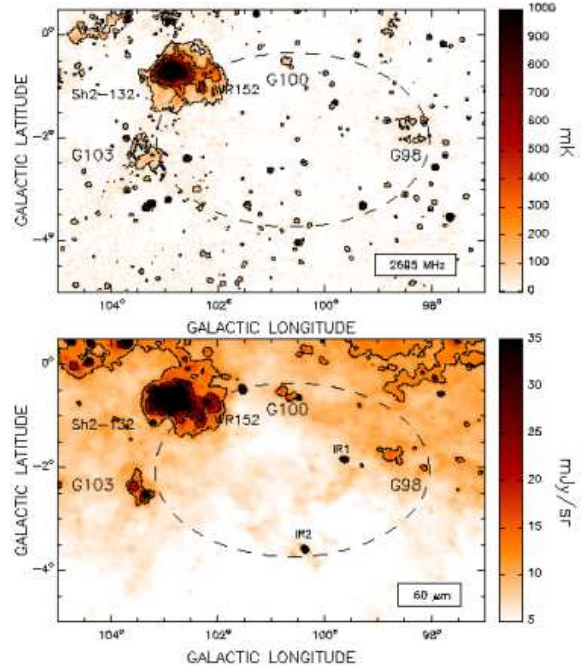}
\caption{{\it Upper  panel}. Effelsberg 2695 MHz emission distribution in the area of \gs.  Contour levels are at  65, 250 and 400 mK. 
{\it Lower panel}. IRIS 60 $\mu$n emission distribution. Contour levels are at  12, 18 and 24 mJy\,sr$^{-1}$.
In both panels  the ellipse marks the location of  the  large \hi\, shell.}
\label{11cm}
\end{figure}

Figure \ref{11cm} shows the observed radio continuum emission at 2695 MHz (upper panel) and the 60 $\mu$m infrared emission 
 (lower panel) towards the region where \gs\, is detected. 

No large scale features that could be interpreted as the counterpart of the \hi\, structure could be found 
neither in the radio continuum nor in the infrared images. 
The strong source observed at both frequencies  at ($l,b$) $\simeq$
 (102\fdg8, --0\fdg7) is the well known \hii\, region Sh2-132 , which has been recently studied by \citet{vas10}. Towards the southwest of this \hii\, region, a ring nebula related to the Wolf-Rayet star WR\,152 is also observed \citep{cap10}.
 On the other hand, three  slightly 
extended and much more weaker sources are seen projected onto the \hi\, shell. 
Following the standard nomenclature, they will be referred to as G103.39-2.28 (G103, for short), G100.7-0.5 (G100),
  and G98.51-1.7 (G98).
These three sources will be  analyzed in Section \ref{sources}.

In the infrared image (see Fig. \ref{11cm} lower panel) two sources,
IRAS 22036+5306 (IR1) and IRAS 22142+5206 (IR2), located at ($l, b$) =
($99\fdg63, -1\fdg85$) and ($l, b$) = ($100\fdg38, -3\fdg58$),
respectively, are observed. Neither of these sources has a counterpart at 2695 MHz. The first source is associated
with a post AGB star, whilst the second one (IRAS 22142+5206) may be a protostellar object \citep{dob98}.

They found  IRAS
22142+5206 to be associated with CO molecular emission at $V = -37.2$ \kms.

\subsection{large \hi\, shell origin.}\label{originob}

The origin of \gs\, may be attributed to the action of the stellar wind of several massive stars and their subsequent supernova explosions. In the following, based on the derived parameters of the shell, we analyze both scenarios.

An inspection of the {\it Galactic OB Associations in the Northern Milky Way Galaxy} presented by \citet{gar92} reveals that several stars 
belonging to Cep\,OB1 are seen  projected inside the ellipse delineating \gs. They are indicated by symbols in Fig. \ref{100-2-37ProV}.

 \citet{gar92} determined for Cep\,OB1 a photometric distance modulus of DM = 12.2 mag, which yields a  distance of 2.75 kpc.
On the other hand,  based on  the stellar proper motions given by Hipparcos, \citet{mel09} estimated  a distance of 2.8 kpc for 
 this  OB association.
 This distance is consistent with the distance 
estimate derived for \gs\, when non-circular motions are considered \citep{bra93}.
To test the possibility that the Cep\,OB1 stars lying inside the \hi\ shell could have been capable of creating it, 
an evaluation of the energy that may be injected into the ISM by the early type stars of Cep\,OB1 is in place. Is the energy provided by these massive stars enough to create \gs?

Those stars lying within the boundaries of \gs, twenty in total, are listed 
in Table \ref{cep}. Column 1 gives the star's identification, Col. 2 and 3  their galactic coordinates, and Col.4  their 
spectral types as given by \citet{gar92}. Based on the evolutionary track models published by \citet{sch92}, and adopting the
 bolometric magnitudes and effective temperatures given by \citet{gar92}, we estimated the main sequence (MS) spectral type 
for each star  (Col. 5). Column 6 gives an estimate of the star MS lifetime ($t$(MS)) as derived from the stellar models of \citet{sch92}. The values 
given in this column are a rough estimate as a consequence of the uncertainty in the mass adopted for each star. 
Column 7 and 8 give the mass loss rates ($\dot{M}$) and wind velocities  ($V_w$) taken from \citet{lei98}, respectively.
Column 9 gives the total wind energy released by each 
star during its main sequence phase, $E_w = 0.5 \,\dot{M}\, V_w^2 \, t (\rm MS)$.   
Given that HDE\,235673 and HDE\,235825 are still in the MS, their values are upper 
limits.

\begin {table*}
\caption{Cep\,OB1 stars lying inside \gs.} 
\label{cep}
\centering
\begin{tabular}{l c c c c c c c c c }
\hline\hline
Star & $ l$ & $b $ &Sp. Type & MS Sp. Type &  MS lifetime (Myr)& log ($\dot{M}$ (M$_{\odot} yr^{-1}$)) & $V_w$ (\kms) & $E_w $($10^{50}$ erg) \\
\hline
HDE\,235673 & 98\fdg4 & --1\fdg6 & O6.5 V & O6.5 & 5.6&  --6.27 & 2700 & $\le$ 2.2\\
HD\,209678 & 99\fdg3 & --1\fdg8 & B2 I& B0 & 12 & --7.17  & 1600 & 0.2\\
HD\,209900 &99\fdg7&--1\fdg7 & A0 Ib & B1 & 26 &  --8.2 & 2500 & 0.2 \\
HD\,210809 &99\fdg9& --3\fdg1 & O9 Iab & O7& 6.4 & --6.39& 2700  & 2.1 &\\
BD\,+51\,3135 & 100\fdg6& --2\fdg1 & B3 II& B0.5 & 14 &  --7.3 & 1850 & 0.2\\
HDE\,235781 &101\fdg2& --2\fdg6 & B6 Ib & B1 & 22 & --8.2 & 2500 & 0.2\\
BD\,+53\,2820 &101\fdg2 & --1\fdg7 & B0 IV & B0 & 16 &  --7.17 & 1600& 0.2\\
BD\,+52\,2833 & 101\fdg4& --2\fdg1 & B1 III & O9.5 & 10 & --7.04 & 2500& 0.5\\
BD\,+53\,2837 & 101\fdg5 & --2\fdg1& B2 III& B0.5 & 14 & --7.3 & 1850 & 0.2\\
HDE\,235783 & 101\fdg7& --1\fdg9 & B1 Ib&  B0.5 & 14 & --7.3 & 1850  & 0.2\\
BD\,+53\,2843 & 101\fdg8 & --2\fdg2 & O8 III & O7 & 6.4 &--6.39 & 2700 & 2.1\\
BD\,+54\,2718 & 102\fdg0 & --0\fdg9 & B2 III & B0.5 & 14 & --7.3 & 1850 & 0.2\\
BD\,+54\,2726 & 102\fdg2 & --1\fdg0 & B1.5 II & B0.5 & 14 & --7.3 & 1850 & 0.2\\
HDE\,235813 & 102\fdg4 & --2\fdg0 & B0 III&  O6 &  6.3 & --6.14 & 2700 & 2.6 \\
HILTNER\,1106 & 102\fdg5 & --0\fdg7 & B0 III & O8 & 6.9 & --6.65 & 2600 & 0.9\\
BD\,+53\,2885 & 102\fdg7 & --2\fdg9 & B2 III & B0.5 & 14 & --7.3 & 1850 & 0.2\\
HD\,212455 & 102\fdg8 & --1\fdg7 & B6 Ib & B0.5 & 14 & --7.3 & 1850  & 0.2\\
HDE\,235825 & 102\fdg9 & --1\fdg8 & O9 V & O9 & 8 & --6.91 & 2500 & $\le$0.6\\
BD\,+54\,2764 & 103\fdg0 & --1\fdg7 & B1 Ib & B0.5 & 14 & --7.3 & 1850 & 0.2 \\
BD\,+54\,2761 & 103\fdg1 & --1\fdg4 & O6 III & O3* & 4.3 & --5.43 & 3200 & $\le$16.3 \\
\hline
\end{tabular}
\tablebib{*: The bolometric magnitude and effective temperature given by \citet{gar92} lie outside the evolutionary tracks of 
\citet{sch92}. We adopted for this star the earliest spectral type.} 
\end{table*}

Theoretical models predict that only 20 \% of the wind energy  is converted into mechanical  
energy of the shell \citep{wea77}. In the case of \gs\,, this implies that for the total kinetic energy stored in the shell, 
$E_k = (1.8 \pm 0.8) \times 10^{50}$ erg, a wind energy greater than 9 $\times 10^{50}$ erg 
would be required. However the analysis of several observed \hi\, shells shows that the energy conversion efficiency seems to be lower, roughly about 2-5 \% \citep{cap03}.

From the last column of Table \ref{cep} we infer that the total wind energy injected during the MS phase of the  Cep\,OB1 stars is 
$\sim 29.7 \times 10^{50}$ erg, which is enough to create \gs\,  if the energy conversion efficiency were $\sim$ 6 \%.

It is important to mention that the 
MS lifetime of all the O stars, which are the main energy contributors,  is compatible, within errors, with the dynamical age of the shell (5.5 $\pm\, 2.7$ Myr). Moreover, the fact that most of the massive stars located inside the shell are evolved stars may explain the absence of ionized gas related to \gs.

On the other hand, taking into account that the SN rate in a typical OB association is about one per $10^5 - 3 \times 10^5$ yr. \citep{mcc87},
over the lifetime of \gs\, is reasonable to assume that the most massive members of Cep\,OB1 may have exploded as SN. In this case, the energy of the explosions, as well as the energy injected by the stellar winds of the SNe progenitors, would have also contributed to the formation of the shell.
There is no detection of a SNR in either radio surveys or  x-ray surveys, but this could be due to the relatively large age of \gs. However, an evidence that a SN explosion could have taken place in the region is the presence of the pulsar PSR\,J2150+5247, located at ($l, b$) = (97\fdg5, --0\fdg9) \citep{tay93}. Even though the estimated distance to the pulsar is 5.48 kpc \citep{tay93}, bearing in mind that this distance is obtained based on the dispersion measure and a galactic electron density model \citep{tay93b}, the error is large, and therefore the possibility that the pulsar were located at a  distance similar to the one of \gs\, can not be ruled out.  
Were this the case, an upper age limit for
 the progenitor of PSR\,J2150+5247, assuming it was an early B-type star, 
would be $\sim$3 $\times$ 10$^7$ yr, that is in reasonable agreement with the age of the oldest star, HD\,209900, belonging to Cep\,OB1.

\section{Sources G98, G100, and G103}\label{sources}

\subsection{Their Nature.}

\begin{figure*}
\centering
\includegraphics[width=17cm]{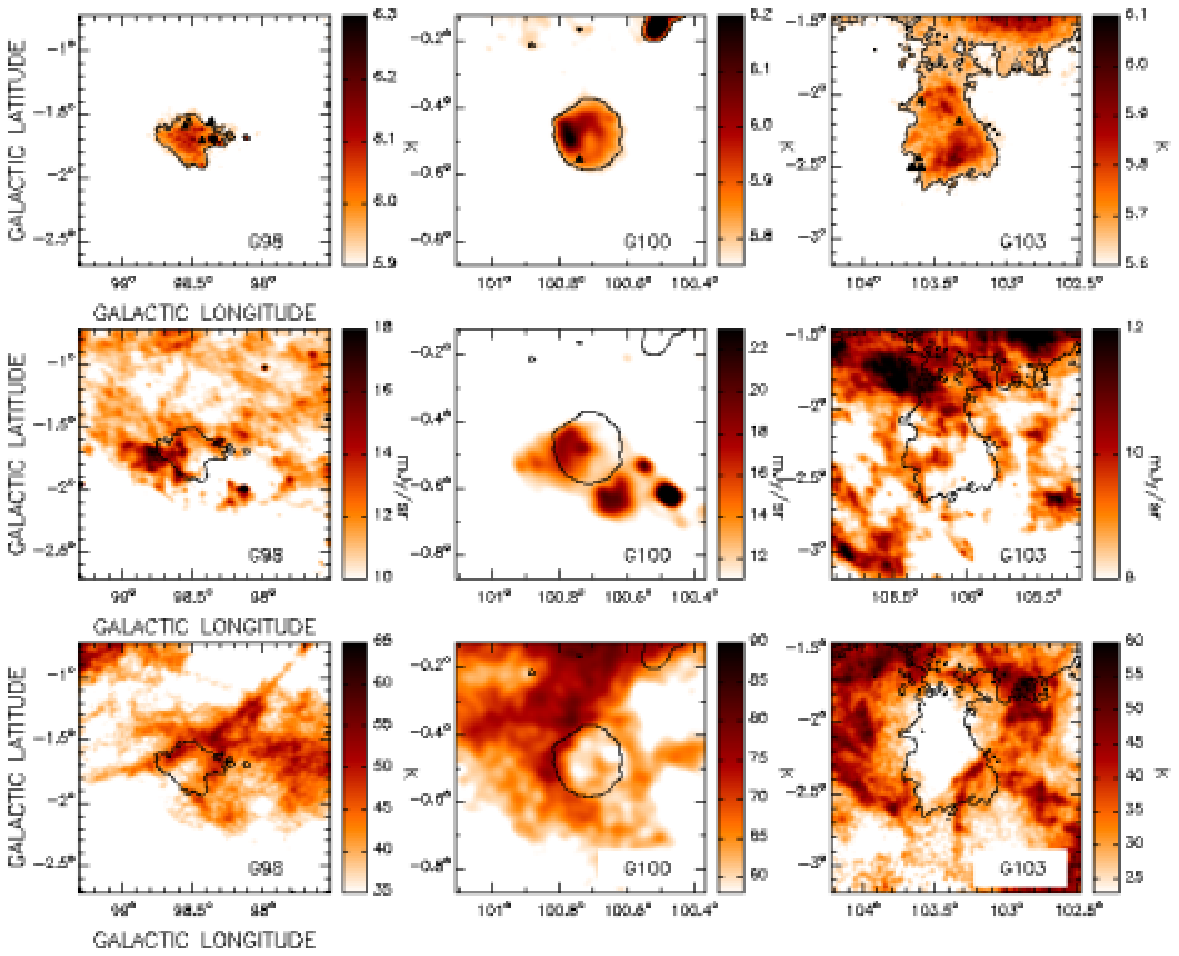}
\caption{{\it Upper panels} 1420 MHz radio continuum images in the area of G98 (left), G100 (middle), and G103 (right).
Point-like sources have been removed. Contour levels are at  6.0, 5.8, and 5.6  K, for G98, G100, and G103, respectively.
The black triangle symbols indicate the location of the stars HD\,235673, ALS\,12073, ALS\,12071 and ALS\,12074 (G98), BD+542684 (G100), and ALS\,12434, ALS\,12443, ALS\,12471, ALS\,12469 and ALS\,12475 (G103).
\textit{Middle panels}: HIRES 60 $\mu$m emission distribution in the area of G98, G100 and G103. For comparison with the 1420 MHz emission, contour levels are the same as in upper panels.
{\it Lower panels}: CGPS \hi\, emission distribution averaged in the velocity range from  --41.9 to -- 50.9 \kms\, in
 the area of G98 (left), from  --50.9 to --55.9 \kms\, in the area of G100 (middle) and from  --29.5 to --31.15 \kms\, in the area of 
G103 (right). Contour levels are the same as for upper and middle panels. In all images the  angular resolution is 2-arcmin.} 
\label{cont}
\end{figure*}

As mentioned in Section \ref{cont-ir}, three  slightly extended sources, labelled G98, G100, and G103, are observed at  
radio continuum and infrared wavelengths, 
projected onto the  border of \gs.
The upper panels of Fig. \ref{cont} show the CGPS 1420
MHz  radio continuum image of each source. In these images, discrete  sources have been removed.  The middle panels  show the HIRES  60 $\mu$m  emission distribution  in the area of G98, G100, and G103. At both fequencies, to increase the signal-to-noise ratio, the original images were smoothed to a 2-arcmin resolution.

A clue onto their physical nature (thermal or non-thermal) may be provided by the behavior of the flux density as a function of
frequency. Previous flux density estimations for G100 were obtained by \citet{whi92} and \citet{ker07}.  \citet{whi92} obtained a flux density at 1420 MHz of 456 mJy, while \citet{ker07} estimated a lower flux density, S$_{1420} = 48.8 \pm 24.0$ mJy.
To estimate the spectral index $\alpha$  ($S_{\nu} \propto \nu^{-\alpha}$)  of each source, CGPS continuum observations at 408 and 1420 MHz, and Effelsberg continuum data at 2695 MHz were used in this work.

Flux density determinations, specially for weak sources like the ones we are dealing with, are strongly dependent on the source angular extent, 
which in turns  depends on the outermost reliable continuum level. 
Before defining the angular extent of each source, the overall galactic continuum emission has to be subtracted out.
To this end we applied the "Back Ground
Filtering" method (BGF) developed by \cite{sof79}. A 1$\degr \times $1$\degr$ filtering beam was applied to all 
sources at each frequency.

The derived flux densities are given in Table \ref{tabla2}. 
The quoted errors stem from the uncertainty in determining the background levels.
The spectral index $\alpha$  was derived from a fit to the observed flux densities, 
under the assumption that none of these sources is optically thick at the lowest frequency (408 MHz). 
For G100 a residual grating ring from the strong nearby Cas\,A source falls on top of the source, making the estimate of the
408 MHz flux density meaningless. The lower flux density obtained by \citet{ker07} is consistent with the smaller area considered for G100 by these authors. On the other hand, the higher value given by \citet{whi92} may be attributed to the fact that they used Green Bank data (HPBW $\sim 11\arcmin $).

Though highly uncertain, the spectral indexes are consistent with a thermal nature for G98, G100, and G103; implying that they are \hii\, regions.
Another argument in favor of this interpretation comes from the far infrared data. From the measured flux densities at 60 and 100 $\mu$m and following the procedure descibed by \citet{cic01}, the dust temperature was estimated for each region (see Table \ref{tabla2}). The derived temperatures are typical of \hii\, regions. The errors quoted for the dust temperatures are formal errors from the fitting procedure.

\begin {table}
\caption{\textbf{G98, G100 and G103 parameters}.} 
\label{tabla2}
\centering
\begin{tabular}{l c c c  }
\hline\hline
Source &G98& G100& G103 \\
\hline
Longitude & 98\fdg5 & 100\fdg7 & 103\fdg3\\
Latitude & --1\fdg6 & --0\fdg5 & --2\fdg4\\
Size (arcmin) &  73 x 64 & 19 x 23 &  54 x 87\\
S$_{408}$ (Jy)&2.1 $\pm$ 0.3& -- &  2.0 $\pm$ 0.3 \\
S$_{1420}$ (Jy)&  2.0 $\pm$ 0.1& 0.21 $\pm$ 0.03 & 1.50 $\pm$ 0.15\\
S$_{2695}$ (Jy)& 1.9 $\pm$ 0.3& 0.19 $\pm$ 0.03 & 1.70 $\pm$ 0.35 \\
Spectral index $\alpha$ &  --0.05 $\pm$ 0.01  & --0.16 $\pm$ 0.33  & --0.11 $\pm$ 0.08\\  
S$_{60}$ (Jy) & 410 $\pm$ 20 & 120 $\pm$ 6  & 1030 $\pm$ 50\\
S$_{100}$ (Jy) & 1640 $\pm$ 80 &  380   $\pm$ 20 & 1840  $\pm$ 90\\
Dust temperature (K)& 24.5 $\pm$ 0.7 & 26.5 $\pm$ 0.8 & 33.1 $\pm$ 1.1 \\
\hline
\end{tabular}
\end{table}

\subsection{Location of G98, G100 and G103.}

We now try to discern whether these ionized regions could be physically related to \gs.
Unfortunately, the radio continuum flux densities of  G98, G100 and G103 are not strong enough to allow us 
to derive a reasonable \hi\, absorption spectrum in order to attempt to set a limit to their distances. Another way to attempt to 
estimate their distances is to look for signatures of the interaction among the ionized regions and  the \hi\, and CO  gas emission observed around these sources. If the continuum sources were related to \gs, the \hii\, regions would expand within the gas of \gs\, and we would be able to observe for each source an \hi\, minimum in a velocity range compatible with the radial velocity of the large \hi\, shell.
After a thorough inspection of the  CGPS \hi\, data cube, we were able to point down \hi\, minima having a good morphological correlation with the radio  continuum sources.

Mean brightness temperature of these structures are shown in the lower panels of Fig. \ref{cont} (the original images were smoothed to a 2-arcmin resolution). The radial velocity ranges where these minima are detected range from --40.2 to --51.8 \kms\,  (G98),  from --50.1 to --58.4 \kms\,  (G100), and from --22.1 to --33.6 \kms\,  (G103). However, it is important to mention that for the averages shown in Fig. \ref{cont} only those velocity channels where the minima are best defined were considered.
To facilitate the comparison among the
radio continuum and the \hi\, emission distributions, the
1420 MHz contour line defining the source's extent is superimposed on the \hi\ images.  

The \hii\, regions are clearly surrounded by enhanced  \hi\, emission. The  positional 
coincidence between the \hi\, features and their corresponding \hii\, region suggests that they are physically related. 
It is worth mentioning that, although the entire velocity cube was searched for \hi\, features likely to be related to the ionized regions, no other peculiar structures but the ones shown in Fig. \ref{cont} were found.

\begin{figure}
\centering
\includegraphics[width=8cm]{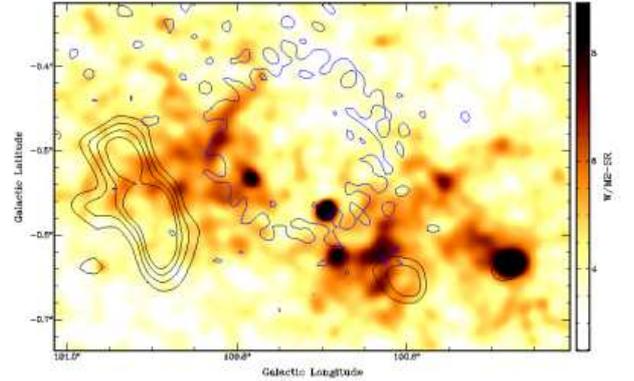}
\caption{8.3 $\mu$m emission distribution in the area of G100. The blue contour correspond to the 5.8 K level at 1420 MHz. Black 
contours are associated with the CO emission distribution averaged between --52.25 and --55.17 \kms, at 0.4, 0.6, 0.8 and 1.0 K.} 
\label{g100}
\end{figure}

We have also analyzed the CO emission distribution in the area of G100
and G103 (unfortunately there is no CO data available for G98). We
found two molecular clouds probably related to G100 in the velocity
range form $-52.25$ to $-55.17$ \kms\, (black contours in Fig. \ref{g100}), in
coincidence with the velocity range where the \hi\, structure was found.
Figure \ref{g100} also shows  the MSX Band A (8.28 $\mu$m) image  (red colours) and the 5.8 K level at 1420 MHz (blue contour).
 Clearly,  the 8.28 $\mu$m emission partially borders the ionized gas.  It is worth
mentioning that the emission observed in the MSX Band A is not
detected in the other three MSX bands (12.13 $\mu$m, 14.65 $\mu$m, and
21.3 $\mu$m), suggesting that the polycyclic aromatic hydrocarbons
(PAHs) could be the main responsible for the emission detected at 8.28
$\mu$m, indicating the presence of a photo-dissociated region (PDR) in
the border of G100.
An inspection of the CO data cube in the area of G103 do not reveal any CO structure probably related to the \hii\, region.

Bearing in mind that \gs\, is detected in the velocity range from $-29$ to $-51.7$  \kms\, and that its baricentral velocity is $V_0 = -41 \pm 2$ \kms, we would locate both G98 and G100 in the  approaching hemisphere of \gs, whilst G103 would be placed in its receding part. In what fallows we adopt for the \hii\, regions the \gs\, distance.

\subsection{Exciting stars of G98, G100, and G103.}

To identify the exciting stars of G98, G100, and G103 we have
inspected the Galactic O Star Catalogue \citep{sot08} and the OB Star
Catalogue \citep{ree03} looking for massive stars located in
projection toward the \hii\ regions. 

Only one star projected within the border of G98 is listed by \citet{sot08}, whilst in the Reed catalogue four stars are seen projected towards G98 (one of them coincides with the one catalogued by \citet{sot08}), one onto G100 , and five onto G103. They are listed in Table \ref{ostars} and their positions are 
indicated by  triangle symbols in Fig.\ref{cont}. It is worth mentioning that HD\,235673, the star seen projected onto G98, appears in Table \ref{cep} as a Cep\,OB1 member and therefore was considered as an input energy source for the creation of \gs. Nonetheless its contribution to the overall energy budget is $\sim$ 7 \%, so if this star were related to G98, its exclusion as being responsible in the creation of \gs\, would not modify the conclusions reached in Section \ref{originob}.

To check whether these stars may have contributed to the creation of
the observed structures,  under the assumption that they are located at the same distance than the \hii\, regions (2.8 $\pm$ 0.6 kpc), we analyze if their absolute magnitudes are compatible with those corresponding to OB-type stars. We need first to estimate the visual absorption in the region.  From an averaged  \hi\, profile (from 0 to --41 \kms) of all over the solid angle covered by \gs, we derive a total \hi\, column density of about $N_{\hi} = 3 \times 10^{21}$ cm$^{-2}$. Using the relation  $A_v = 5.3 \times 10^{-22} N_{\hi}$ \citep{boh78}, we estimate $A_v = 1.6$ mag. As a check on this, for  HD\,235673  we estimated the visual absorption using  $A_v = 3.1\, E(B-V)$, where $E(B-V) = (B-V)- (B-V)_0 $. For $(B-V)_0 = -0.32$  \citep{sch82} and $(B-V)= 0.21$ \citep{hil56} we   obtain $A_v = 1.6$. Both values are in agreement with the absorption of 1-2 mag given by \citet{nec80}.Then, adopting a visual absorption of 1.6 mag, the absolute magnitudes (see Column 5) of the stars listed in Table \ref{ostars} were calculated.  Based on  \citet{mar06}  and  \citet{sch82}, the corresponding spectral types were estimated (see Col. 6 of Table \ref{ostars}). 

Given that at the assumed distance all the stars listed in Table \ref{ostars} could be O- or early B-type stars, we
suggest that they may be 
responsible of creating G98, G100, and G103.

\begin {table*}
\caption{OB stars probably related to G98, G100, and G103.} 
\label{ostars}
\centering
\begin{tabular}{l c c l c c  }
\hline\hline
Star & Galactic coordinates ($l, b$)&  Sp. Type & v (mag) & M$_v^{(a)}$ & Sp. Type$^{(a)}$  \\
\hline
\bf{G98} & & & & & \\
HD\,235673$^{(b)}$ & 98\fdg36, --1\fdg55 & O7  & 9.14 & --4.7 $\pm$ 0.5& O5/O8.5\\
ALS\,12073 & 98\fdg43, --1\fdg71 & OB   &12.1 & --1.7  $\pm$ 0.5& B2/B5 \\
ALS\,12071 & 98\fdg53, --1\fdg55 & OB & 12.25 & --1.6 $\pm$ 0.5 & B2/B5 \\
ALS\,12074 & 98\fdg55, --1\fdg59 & OB & 12.04 & --1.8 $\pm$ 0.5 & B2/B5 \\
\bf{G100} & & & & &  \\
BD\,+542684 & 100\fdg74, --0\fdg55 & OB & 10.8 & --3.0 $\pm$ 0.5& B0/B2 \\
\bf{G103} & & & & & \\
ALS\,12434 & 103\fdg32,  --2\fdg18 & B5 & 10.6 & --3.2 $\pm$ 0.5 & B0/B2 \\
ALS\,12443 & 103\fdg59, --2\fdg04 & OB & 12.6 & --1.2 $\pm$ 0.5 & B3/B7 \\
ALS\,12471 & 103\fdg59, --2\fdg50 & OB & 12.4 & --1.4  $\pm$ 0.5 & B2/B7\\
ALS\,12469 & 103\fdg61, --2\fdg44 & B2 & 10.6 & --3.3  $\pm$ 0.5 & B0/B2 \\
ALS\,12475 & 103\fdg66, --2\fdg50 & B2 & 11.4 & --2.4  $\pm$ 0.5 & B1/B3\\
\hline
\end{tabular}
\tablebib{(a): For the adopted distance of 2.8 $\pm$ 0.6 kpc. (b): Star found in both catalogues and member of Cep\,OB1 \citep{gar92}.}
\end{table*}

Next, we should consider whether the number of UV ionizing photons
needed to keep the continuum sources ionized could be provided by their
probably exciting stars.  The total number of Lyman continuum photons is given by  N$_{Lym}= 0.76 \times 10^{47}\,
$T$_4^{-0.45}\, \nu_{\rm GHz}^{0.1} $D$_{\rm kpc}^2$ \,S$_\nu$ \citep{cha76}, where
T$_4$ is the electron temperature in units of $10^4$ K, D$_{\rm kpc}$
is the distance in kpc, $\nu_{\rm GHz}$ is the frequency in GHz and
S$_\nu$ is the measure flux density in Jy.  Using the 1420 MHz flux
densities given in Table \ref{tabla2}, a distance of 2.8 kpc and
adopting T$_4$ = 1, we obtained N$_{Lym} = 1.2 \times 10^{48}$
s$^{-1}$ for G98, N$_{Lym}= 1.3 \times 10^{47}$
s$^{-1}$ for G100, and N$_{Lym}= 1.8 \times 10^{48}$
s$^{-1}$ for G103. According to the theoretical models of \citet{sch97}, for solar metallicity, the estimated N$_{Lym}$
 necessary to keep G98, G100, and
G103 ionized, could be provided by a O9.5V, B0.5V, and O9.5V stars (or two B0V), respectively.  This indicates that the join action of the stars listed in Table \ref{ostars} 
 can provide the ionizing photons needed for each 
\hii\ region. In the case of G98, given that the contribution of HD\,235673 is essential to keep this region ionized, 
we conclude that this star is associated with G98 instead of being related to the genesis of \gs\, as it was considered in Section \ref{originob}. This implies that the wind energy provided for this star should not have been taken into account for the origin of \gs. 
As mentioned before, this fact does not affect the scenario proposed for the creation of the large \hi\, shell.

\section{Triggered star formation?}

It is generally believed that expanding shells may induce star
formation at their edges \citep{elm98}. Shells behind shock fronts
experience gravitational instabilities that may lead to the formation
of large condensation inside the swept-up gas, and some of them may
produce new stars. An increasing body of observational evidence
supports the importance of the shell's evolution in creating new
stars \citep{pat98, oey05, arn07, cic09, cic11}. 

 In our case, having
an old large \hi\, shell containing several \hii\, regions in its edge that
seems to be at the same distance, we wonder if this could be other
case of triggered star formation.  Were this the case, we should expect an age gradient in the region, in the sense that G98, G100, and G103 should be younger than \gs.
In what follows we shall attempt to
estimate the age of the \hii\, regions G98, G100, and G103, and
compare them with the dynamical age of \gs\, estimated in Section
\ref{hi} (see Table \ref{tabla1}). 

As a rough estimate to the age of each region, we evaluate their dynamical ages as $t_{dyn} = 0.6 \, R/ V_{e}$ \citep{wea77}.
The radius $R$ of each region was estimated from the sizes given in Table \ref{tabla2}. We infer $R = 27.8\, \pm\, 6.0$, $8.5\, \pm\, 1.8$ and $27.9\, \pm\, 6.0$ pc, for G98, G100 and G103, respectively, where the errors stem from the uncertainty in the distance.  
Relying on the velocity ranges were the \hi\, emission associated with each \hii\, region is observed (see Section \ref{sources}),  expansion velocities of 5.8 \kms\, (G98), 4.2 \kms\, (G100) and 5.8 \kms\, (G103) are assumed. Thus, we 
 obtain an age of 2.9 $\pm\, 0.9$, 1.2 $\pm\,  0.5$ and 2.9 $\pm\, 0.9$ Myr for G98, G100,
and G103, respectively. The quoted errors stem from the uncertainties in the sizes and from the assumption that  the expansion velocities are accurate to within 1.3 \kms\, (one velocity channel).

From these estimates,  we conclude that  G98, G100, and G103 are younger than \gs.
This age difference is supported by the fact  that 
 most of the stars probably responsible of creating \gs\, have already evolved from the MS, while the ionized gas of the \hii\, 
regions is still observed.

As mentioned in Section \ref{cont-ir}, in addition to G98, G100, and G103, three interesting sources are seen projected onto the 
border of \gs: the \hii\, region Sh2-132, the ring nebula associated with WR\,152, and the infrared source IRAS\,22142+5206.

Regarding Sh2-132 and the ring nebula associated with WR\,152,  \citet{vas10} and \citet{cap10} found neutral gas interacting with the \hii\ regions in the velocity ranges from
--38 to --53 \kms\, and from --43 to --52 \kms\,, respectively. For Sh2-132 this is in coincidence with the ionized velocity gas \citep{geo76, chu81, rey88, fic90, qui06b}. Taking into account non-circular motions in this part of the Galaxy \citep{bra93}, \citet{vas10} and \citet{cap10} inferred for these regions a kinematical distance of $3.5 \pm 1.0$ kpc. They argued that this value is in close agreement with the distance estimates of the main exiting stars of the regions, WR\,153ab and WR\,152. On the other hand, bearing in mind that \gs\ is observed in the velocity range from --29.0 to --51.7 \kms\,, and that the \hii\, regions are located onto its border, we suggest that both ionized structures may be 
located at the same distance as \gs, and that the velocity differences are because the \hii\, regions are located in the approaching part of the large shell. 
Although \citet{vas10} do not give any age estimate for Sh2-132, the fact that it is being ionized by
WR\,153ab, implies that Sh2-132 should be older than the period of time spent by the progenitor of the WR in the main sequence phase. 
Given that WR\,153ab is a  WN6.5 star, the mass of the progenitor was probably about 40-50 $M_{\odot}$ \citep{cro07}.
According to \citet{sch92} the MS lifetime for such a star is about 4 Myr. Concerning the age of the ring nebula associated with  WR\,152, \citet{cap10} estimated a dynamical age of  1 Myr for the associated wind blown bubble. However, knowing that, as mentioned by \citet{cap10}, large errors are involved in this estimate and bearing in mind the time that the progenitor of the WR star spent in the MS phase, the age of the \hii\, region is probably higher, closer to the one estimated above for Sh2-132.

With respect to IRAS\,22142+5206, as mentioned in Section \ref{cont-ir}, this infrared source has a massive molecular 
outflow associated with it at the velocity of --37.2 \kms\, \citep{dob98}. The coincidence between this velocity and
 the velocity range where \gs\, 
is observed suggests they may be located at the same distance. 
Based on the total luminosity of IRAS\,22142+5206, \citet{dob98} suggest that this source will evolve into a late O-type or an 
early B-type star.

In summary, the fact that G98, G100, G103, Sh2-132, the ring nebula associated with WR\,152, and IRAS\,22142+5206 lie at the edge of \gs\, and seems to be at the same 
distance, together with the age gradient, suggest that these sources could have been triggered by the expansion of \gs.

\section{Conclusions}

The large \hi\, shell \gs\, has been analyzed to study the interaction of massive stars with the interstellar medium and, in 
particular, the process of triggered star formation. 
From this analysis we conclude the following:

\begin{enumerate}

\item \gs\, is a large shell of a radius of about 102 pc located at a distance of 2.8 $\pm\,$ 0.6 kpc. The swept up mass in the shell is ($1.5 \pm\, 0.7$) $ \times\, 10^5$ M$_\odot$ and the shell density $n_{sh} = 2.5 \pm\, 0.4$ cm$^{-3}$. The shell is expanding at a velocity of 11 $\pm\, 2$ \kms\, and its kinetic energy is ($1.8 \pm\, 0.8$) $\times\, 10^{50}$ erg.

\item Several evolved massive stars members of Cep\,OB1 are projected inside the large shell. The distance to the OB association is compatible with the kinematical distance of \gs\, when non-circular motions are considered. An energetic analysis suggests that the wind energy provided during the main sequence phase of the stars could explain the origin of the shell. However, taking into account the SN rate in OB associations, the energy contribution of a SN explosion as well as of its massive progenitor can not be discarded. 

\item From the 2695 MHz radio continuum and 60 $\mu$m infrared images, we found three slightly extended sources, labelled  G98, G100, and G103 projected onto the borders of \gs. From the radio flux densities estimated at different wavelengths, the thermal nature of the sources was confirmed by the estimation of their spectral indexes. In addition, dust temperatures were estimated and found to be typical of \hii\, regions.

\item An inspection of the 1\am\, CGPS \hi\, data reveals \hi\, minima  having a good morphological correlation with the \hii\, regions at velocity ranges compatible with the velocity spanning by \gs.  This leads to the conclusion that G98, G100, and G103 are located at the same distance than \gs.

\item  From O and OB star catalogues, the massive star candidates to be responsible for the ionized gas were identified. 

\item The obtained age difference among the \hii\, regions and the shell, together with their relative location leads us to the conclusion that G98, G100, and G103 may have been created as a consequence of the action of a strong shock produced by the expansion of \gs\, into the surrounding gas.

\end{enumerate}

\begin{acknowledgements} 

The CGPS is a Canadian Project with international partners and is supported by grants from NSERC. Data from the CGPS are publicly available through the facilities of the Canadian Astronomy Data Centre (http://cadc.hia.nrc.ca) operated by the Herzberg Institute of Astrophysics, NRC. This project was partially financed by the Consejo Nacional de Investigaciones Cient\'{i}ficas y T\'ecnicas (CONICET) of Argentina under project PIP 01299, Agencia PICT 00902, UBACyT 20020090200039, and UNLP G091. L.A.S is a doctoral fellow of CONICET, Argentina. S.C and M.A. are members of the \textit{Carrera del Investigador Cient\'{i}fico} of CONICET, Argentina. J.C.T. is member of the \textit{Carrera del Personal de Apoyo}, CONICET, Argentina.

 \end{acknowledgements}

\bibliographystyle{aa} 
\bibliography{bibliografia}
  
 \IfFileExists{\jobname.bbl}{}
{\typeout{}
\typeout{****************************************************}
\typeout{****************************************************}
\typeout{** Please run "bibtex \jobname" to optain}
\typeout{** the bibliography and then re-run LaTeX}
\typeout{** twice to fix the references!}
\typeout{****************************************************}
\typeout{****************************************************}
\typeout{}
}

\end{document}